\documentclass[twocolumn,prl]{revtex4}

\usepackage{epsfig}
\usepackage{amsmath}
\usepackage{graphicx}
\usepackage{dcolumn}
\usepackage{amsmath}
\usepackage{latexsym}

\begin{document}

\title{Two gate-voltage periods in a metallic-nanoparticle based single-electron transistor}
\author{L. Bitton}
\author{R. Berkovits}
\author{A. Frydman*}
\address{The institute of nanotechnology and advanced materials, The Department of Physics, Bar Ilan University, Ramat Gan 52900, Israel}

\begin{abstract}

Systems of quantum dots (QD) connected to leads exhibit periodic conductance peaks as a function of gate voltage arising from the Coulomb blockade effect \cite{review1,review2,review3}.
Much effort goes into minimizing the size of QDs and reaching the scale of single molecules
\cite{molecular1,molecular2,molecular3} which could serve as nanoelectronic circuit components such as transistors. Connecting molecules or nanoparticles to external leads cannot be achieved by the traditional methods used in semiconductor based
QDs, hence, controlling the coupling to nanoparticle QDs is a major technical challenge.
 Here we present a novel technique by which we can explore electric properties of a metallic nanoparticle while
 varying the coupling to leads. We find that the conductance through the nanoparticle is characterized
  by \emph{two} gate voltage periods. The relative strength of the periods depends both on the dot-lead coupling and
  on the source-drain voltage. These surprising findings may be a  general property of strongly coupled metallic
  nanoparticles.
\end{abstract}
\maketitle

 The electronic properties of a dot-lead system are governed by four energy scales, $k_{B}T$ - the thermal energy,
 $E_{C}$ - the charging energy, $\Delta$ - the single electron energy level spacing and $\Gamma$ - the level width
 that depends on coupling to the leads. The ratio between  $\Gamma$ and $\Delta$ distinguishes between weakly-coupled
 and strongly-coupled dots. Weak coupling ($\Gamma << \Delta$ corresponding to conductance, $\sigma \ll e^{2}/h$) is
 characterized by well defined Coulomb blockade (CB) peaks, each peak representing a single electron filling of a consecutive single
  electronic level. 
  \begin{figure}[h!b]

 {\epsfxsize=3.4 in \epsffile{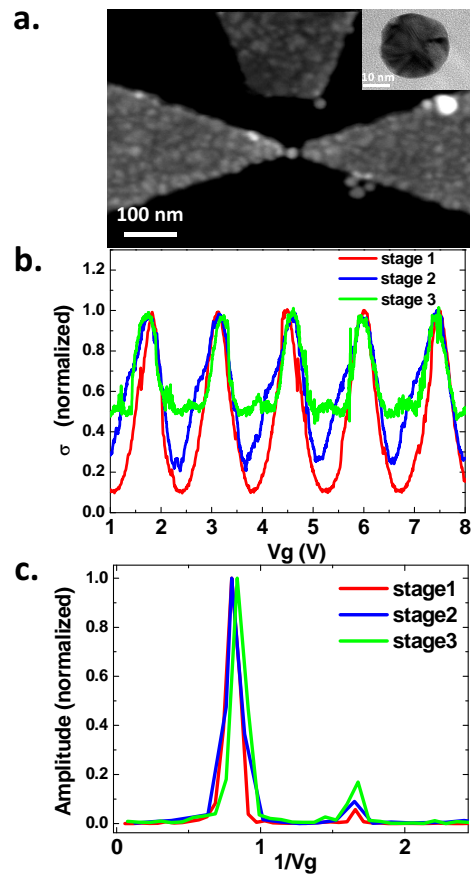}}
 \vspace{-0.6cm}
 \caption{\textbf{a.} Scanning electron microscopy (SEM) image of our dot-leads system. A 30nm gold colloid is placed between the source and drain and a side gate electrode is fabricated 150 nm away. Inset: Transmission electron microscopy image of the colloid.  \textbf{b.} Conductance as a function of gate voltage for three different coupling strengths taken at T=4.2k and $V_{SD}=1mv$. \textbf{c.} Fourier Transform analysis of the three coupling stages, normalized to the slow period amplitude. \small}
\end{figure}
  CB in the strong-coupling limit,
 ($\Gamma \sim \Delta$, $\sigma \sim e^{2}/h$), is expected to be considerably suppressed \cite{matveev, shoen} and
 the transport should be determined by mesoscopic phenomena such as quantum interference \cite{ucf}. Recently,
 exotic CB effects have been predicted for the case where few levels are strongly coupled to the leads
 while all others are weakly coupled \cite{weidenmuller, silvestrov}. Population switching between levels may occur leading to nontrivial asymmetric conductance peaks \cite{silvestrov,richard}.

\begin{figure*}
{\epsfxsize=7 in \epsffile{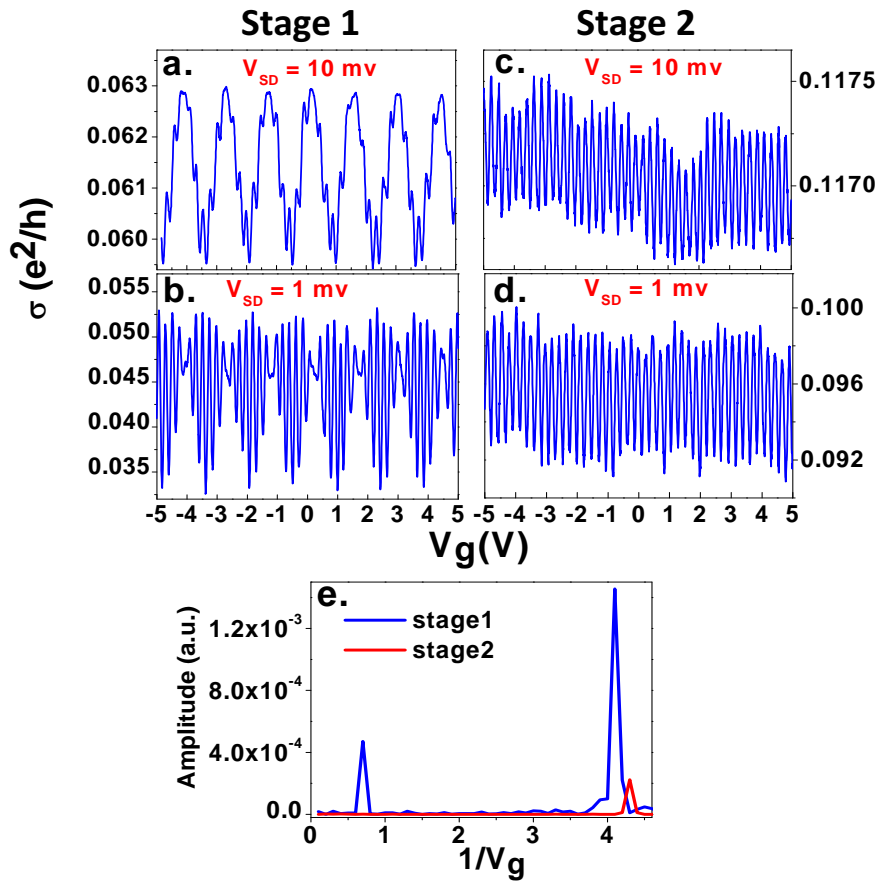}}
\vspace{-0.6cm}
\caption{
 \textbf{a-d.} Conductance vrs
$Vg$ for two different coupling degrees and two values of bias
voltage of one of our quantum-dots. \textbf{e.} Fourier analysis of
the conductance measurements of the two coupling stages taken at
$V_{sd}=10mv$. \small}
\end{figure*}

In order to explore the transition from closed to open dots one would like to be able to control the dot-lead coupling.
In structures based on low-density 2D electron gases, coupling can be controlled by applying back gate
voltages. A similar technique for controlling the coupling between leads and a high-density, metallic QDs doesn't exist and varying the coupling  presents a major challenge. Nevertheless, variable coupling in
metallic QDs may be of great interest since $\Delta$  is orders of magnitude
smaller than $E_{C}$ . This provides a large range for varying
$\Gamma$ through $\Delta$ while still keeping it smaller than
$E_{C}$.  Several methods have been used for studying QDs based
on metallic nanoparticles or other nano-objects. These include discontinuous films \cite{ralph1},
electromigration \cite{park}, electrostatic trapping \cite{trapping, kuemmeth},  break-junctions
\cite{breakjunctions} and angle evaporation \cite{klein1}. None of these techniques provide a way to
control $\Gamma$ and these dots are usually very weakly coupled. We have developed a novel
technique for fabricating a metallic nanoparticle based QD while controllably varying the coupling to electrodes.

Our dots are composed of gold colloids having diameters of 30nm. This yields $\Delta\sim 10 \mu V$  and $E_{c}\sim 25 meV$. Coupling to leads is achieved as follows \cite{liora} : On a Si-SiO substrate we fabricate two gold electrodes (source and drain) separated by a gap of 10-30nm and a perpendicular side gate electrode at a distance of 150nm as shown in fig. 1. We then deposit an adhesive layer of Poly-L-Lysine on the substrate and spread gold colloids on top. Next, we use Atomic Force Microscope (AFM) nano-manipulation to "push" a desired colloid to between the source and drain electrodes. At this stage the dot is usually very weakly connected to the leads. We vary the coupling by depositing gold atoms on top of the gold electrodes using an electrodeposition method. The substrate is placed in a solution containing potassium cyanaurate, potassium bicarbonate and potassium hydroxide \cite{marcus}. A deposition current of $1\mu A$ is applied between the working electrode (for which we use the source and drain) and a gold counter electrode placed in the solution. This results in atomic gold growth on the two electrodes with extremely fine controllable rate. During the deposition process we measure the conductance between the source and drain and stop the process at any desired coupling. We then cool the system to T=4.2K and measure transport properties. We can further apply the electrodeposition process to continue increasing the coupling strength, thus scanning the entire regime from weak to strong coupling.

Conductance versus gate voltage, $\sigma(Vg)$ of a
typical QD system is shown in fig. 1b for three coupling strengths. Clear periodic conductance peaks, are observed for all stages. As coupling is increased two trends become apparent.  First, the amplitude of the peaks decreases with coupling, as expected from the strongly coupled regime. We note that we can not directly estimate $\Gamma$ from the conductivity because our dots might be highly asymmetric. $\sigma$ is governed by the less connected lead, thus, the total system conductance may be much smaller than $e^2/h$ but the coupling to the well connected lead may be relatively high. For this reason we use the visibility $\Lambda\equiv \sigma(peak)/\sigma(dip)$ as a parameter to compare coupling strengths of different samples. The second trend is that, as coupling becomes larger, additional
structure emerges. The weakly coupled dot (red curve) exhibits symmetric conductance peaks. Increasing the coupling
gives rise to asymmetric peaks (blue curve) which eventually take the form of a double-peak structure (green curve).
The  Fourier transform (2b) reveals that for the three coupling stages $\sigma(Vg)$ is composed of \emph{two} periods
and that the relative amplitude of the faster period increases with increasing coupling.

Two periods were observed in all quantum dots (over 10) that had a resistance between 0.5-5 $M\Omega$. A striking example
is shown in fig. 2 which depicts $\sigma(Vg)$ for two coupling strengths and two values of source-drain voltage. It is
seen that for relatively weak coupling and large bias voltage ($V_{SD}=10mv$), $\sigma(Vg)$ is dominated by a slow
period with a small superimposed sixth harmonic (panel 2a). As coupling is increased (panel 2c), a faster period takes
over. Though this fast period is close to the harmonic observed in panel 2a, Fourier analysis (panel 2e) reveals that
it is not an exact harmonic of the slow period. Fig. 2 also demonstrates that a similar change in conductance
oscillation period can be achieved by varying the source-drain bias voltage. For $V_{SD}=10mv$ of stage 1 (2a),
$\sigma(Vg)$ is dominated by a slow period , while at $V_{SD}=1mV$ (2b), a fast period governs . Hence, a transition
from a slow to a fast period can be achieved either by increasing the coupling or by deceasing the
bias voltage.

\begin{figure}
{\epsfxsize=5.5 in \epsffile{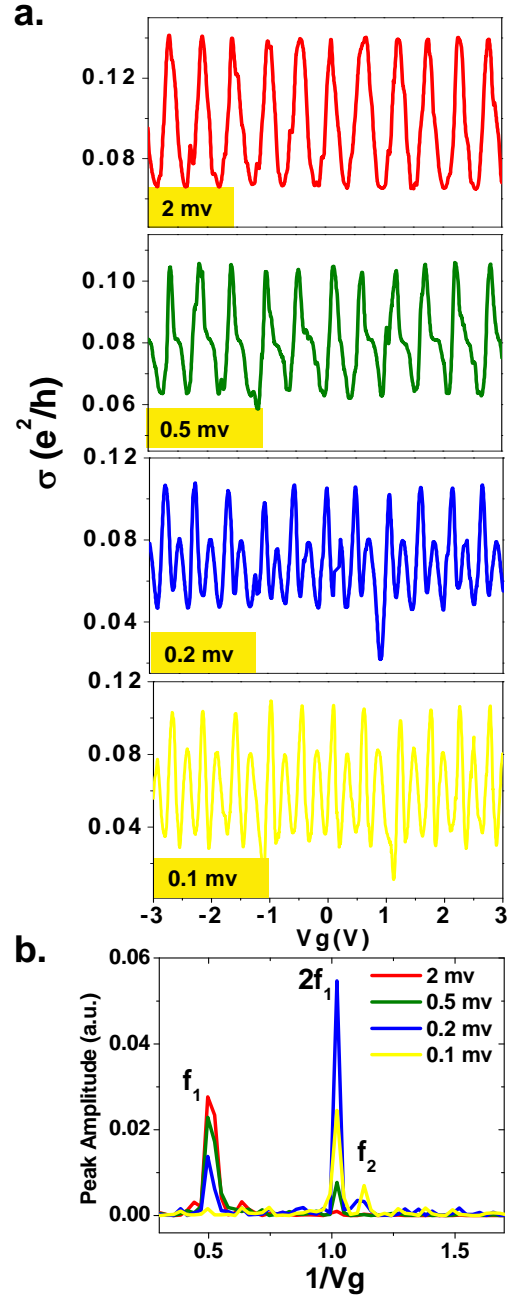}} \vspace{-2cm}
\caption{{a.} Conductance versus gate voltage for an Au dot at
different values of source-drain voltage.  \textbf{b.} Fourier
analysis of each of the conductance measurements. \small}
\end{figure}
Fig. 2 depicts a dot in which the fast period is close to the sixth harmonic of the original period. In the rest of our
dots the fast period observed at large coupling or small $V_{SD}$ was close to twice the slow one. Fig. 3 shows
$\sigma(Vg)$ of such a dot for which the dependence of the conductance peaks on $V_{SD}$ is shown in detail. It is
seen that for large $V_{SD}$ (2mv), $\sigma(Vg)$ exhibits a relatively slow periodic curve characterized by a well
defined frequency, $f_{1}$. As $V_{SD}$ is reduced, an additional frequency, $2f_{1}$, emerges and becomes larger in
amplitude than the $f_{1}$ peak. Further decreasing $V_{SD}$ results in amplitude decrease of the $2f_{1}$ period and
the appearance of a new frequency, $f_{2}$,  that is close to $2f_{1}$ but is not an exact harmonic.

The above findings can be summarized by the following trend for the period dominating the conductance peaks as $V_{SD}$
is decreased or coupling increased: \textbf{$f_{1}\mapsto f_{1}+n*f_{1}\mapsto f_{2}$} with $f_{2}\sim n*f_{1}$ and n
observed between 2 and 6. It turns out that the value of $V_{SD}$ below which $f_{2}$ starts playing a significant
role depends on the coupling. This is demonstrated by comparing figures 2 and 3. The dot of fig. 3 is a relatively
weakly coupled ($\Lambda=2$ at $V_{SD}=1mV$). In this sample the faster period becomes measurable only below 1mV.
In stage 1 of fig. 2 (panels 2a and 2b), $\Lambda_{1mV}=1.5$ and the fast period is observed for $V_{SD}$ as high as
10mV and for stage 2 of this dot (2c and 2d), which has the
strongest coupling, $\Lambda=1.08$, the fast period is observed for
the entire range of measured bias voltages $V_{SD}\leq25mV$. It
appears, therefore, that the emergence of the fast period depends on
the interplay between $\Gamma$ and $V_{SD}$.

The observation of two conductance periods in a quantum dot is a very unexpected result. A natural cause for such an
effect would be the presence of two dots participating in the transport. This seems very unlikely in our case for a
number of reasons. First, we did not observe an additional particle between the electrodes at any  stage of the fabrication process
using SEM or AFM imaging. Usually, the dots should be about a factor of two different in size. Hence, the two dots should be easily observable using advanced microscopy.  Furthermore, before ''pushing'' a gold particle between the source
and drain we never measured peaks in the $\sigma(Vg)$ curves. We checked the possibility that a second dot is created
during the electrodeposition process by growing our electrodes one towards the other without placing a dot in the gap.
These samples exhibited featureless $\sigma(Vg)$.  Chemical analysis showed that the gold atoms are evaporated only
on the electrodes and not on the SiO substrate. It is hard, therefore, to see what would be the origin of a second dot
in our system and we are compelled to assume that the cause for the two periods is an inherent property of the single
     nanoparticle.

\begin{figure}
{\epsfxsize=6.5 in \epsffile{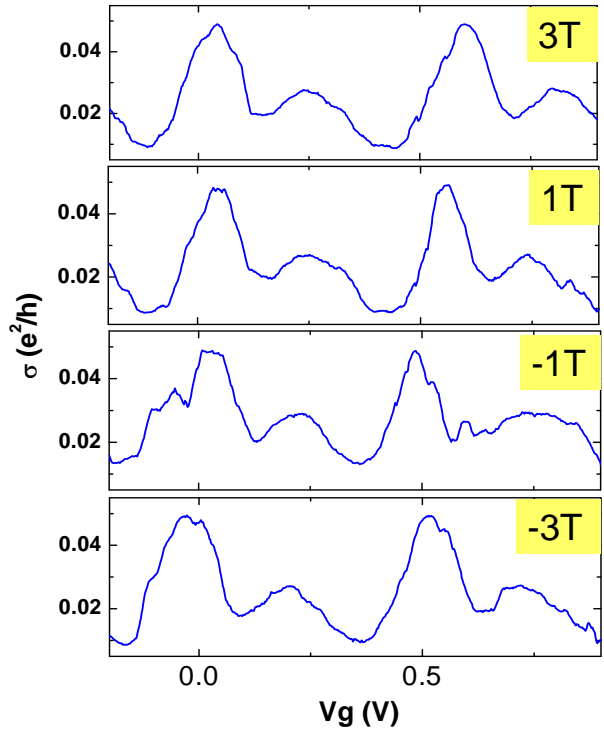}}
\vspace{-1.2cm}
\caption {Conductance as a function of gate voltage at different values of
applied magnetic field. \small}
\end{figure}
We have considered the possibility that the fast period is not a CB effect but rather a result of quantum
interference  between trajectories in which $Vg$ changes the phase difference between the electronic trajectories via
$\Delta k_{F} \cdot l$.  Here $l$ is the path length difference between a couple of electronic trajectories dominating the
transport through the dot. The value of $l$ which corresponds to the measured
conductance periodicity and $\Delta k_{F}$ is a few tens of nm, which is about the size of the dot. This should manifest itself by Aharonov-Bohm conductance oscillations as a function of
magnetic field, H. The magnetic field period that corresponds to the area enclosed by the trajectories is $\sim6T$.
We have performed $\sigma(Vg)$ measurements at various magnetic fields in the range of $\pm 3T$ parallel to the
substrates on three of our dots and have not been able to detect any significant magnetic field affect. Fig. 4 shows
an example for such measurements in which the curves at different fields are identical except for typical scan to scan
variations. Thus, unless the electronic trajectories in the dot are confined to the substrate plane,
quantum interference can not be invoked in order to explain the periodicity of the conductance.

An alternative explanation for switching between two periods may be the fact that typical nanoparticles are chemically formed and are coated by a protective molecular layer of organic compounds and salts. It has been suggested \cite{surface_states} that these may give rise to electronic surface states creating an excess charge plane above the dot. Hence, the electronic states may be divided into surface and bulk states which may play the role of a "two dot" system. Population switching between the two sets of states may result in two conductance-peak periods, inherent to a strongly coupled nanoparticle based QD, although the application of such a scenario to explain our findings would require a better theoretical understanding of population switching as a function of coupling, source-drain voltage and the interplay between them.

\vspace{1cm}

\textbf{Acknowledgements:} This research was supported by the Israeli academy of science (grant number 249/05).
We are grateful for fruitful discussions with Y. Gefen, M. Goldstein, D. Golosov, E. Kogan and M.E. Raikh.

*email: aviad.frydman@gmail.com


\begin{references}
\bibitem{review1} Kastner, M. A. \textrm{The single-electron transistor.} \emph{Rev. Mod. Phys.} \textbf{64,} 849-858 (1992).
\bibitem{review2} Meirav, U. $\&$ Foxman,  E. B. \textrm{ Single-electron phenomena in semiconductors.}\emph{ Semicond. Sci. Technol.} \textbf{11,} 255-284 (1996).
\bibitem{review3} Alhassid, Y. \textrm{The statistical theory of quantum dots.}\emph{ Rev. Mod. Phys.} \textbf{72,} 895-968 (2000)
\bibitem{molecular1} Joachim, C., Gimzewski, J. K. $\&$ Aviram, A. \textrm{Electronics using hybrid-molecular and mono-molecular devices.}\emph{ Nature} \textbf{408,} 541-548 (2000).
\bibitem{molecular2} Park, J. et al. \textrm{Coulomb blockade and the Kondo effect in single-atom transistors. }\emph{Nature} \textbf{417,} 722-725 (2002).
\bibitem{molecular3} Smith, R.H.M. et al. \textrm{Measurements of the conductance of a hydrogen molecule.} \emph{Nature} \textbf{419,} 906-909 (2002).

\bibitem{matveev} Furusaki, A. $\&$ Matveev K. A. \textrm{Coulomb Blockade Oscillations of Conductance in the Regime of Strong Tunneling. }
\emph{Phys. Rev. Lett.} \textbf{75,} 709-712 (1995).
\bibitem{shoen} Herrero, C. P., Schon, G. $\&$ and Zaikin, A. D. \textrm{Strong charge fluctuations in the single-electron box: A quantum Monte Carlo analysis. } \emph{ Phys. Rev. B} \textbf{59}, 5728-5737 (1999).
\bibitem{ucf} Bird, J. P. et al. \textrm{Interference and interactions in open quantum dots.} \emph{Rep. Prog. Phys.} \textbf{66,} 583-632 (2003).
\bibitem{weidenmuller} Hackenbroich, G., Heiss, W. D. $\&$ Weidenmuller, H. A. \textrm{Deformation of Quantum Dots in the Coulomb Blockade Regime.}
\emph{Phys. Rev. Lett.} \textbf{79,} 127-130 (1997).
\bibitem{silvestrov} Silvestrov, P.G. $\&$ Imry, Y. \textrm{Towards an Explanation of the Mesoscopic Double-Slit Experiment: A New Model for Charging of a Quantum Dot.} \emph{Phys. Rev. Lett.} \textbf{85,} 2565-2568 (2000).
\bibitem{richard} Berkovits, R., von Oppen, F. $\&$  Gefen, Y. \textrm{Theory of Charge Sensing in Quantum-Dot Structures.} \emph{Phys. Rev. Lett. }\textbf{94,} 076802 (2005).
\bibitem{oreg} Sindel, M., Silva, A., Oreg, Y.  $\&$ von Delft, J. \textrm{Charge oscillations in quantum dots: Renormalization group and Hartree method calculations.} \emph{Phys. Rev. B} \textbf{72,} 125316(2005).
\bibitem{ralph1} Ralph, D. C., Black, C. T. $\&$ Tinkham, M. \textrm{Spectroscopic Measurements of Discrete Electronic States in Single Metal Particles. } \emph{Phys. Rev. Lett.} \textbf{74,} 3241-3244 (1995).
\bibitem{park} Park, H., Lim, A. K. L., Alivisatos, A. P., Park, J.  $\&$ McEuen, P. L. \textrm{Fabrication of metallic electrodes with nanometer separation by electromigration.} \emph{Appl. Phys. Lett.} \textbf{75,} 301-303 (1999).
\bibitem{trapping} Bezryadin, A., Dekker, C., Schmid, G. \textrm{Electrostatic trapping of single conducting nanoparticles between nanoelectrodes.} \emph{Appl. Phys. Lett} \textbf{71,} 1273 (1997).
\bibitem{kuemmeth} Kuemmeth, F., Bolotin, K. I., Shi, S. F. $\&$ Ralph, D. C. \textrm{Measurement of discrete energy-level spectra in individual chemically-synthesized gold nanoparticles.} \emph{Nano Letters} \textbf{8,} (No.12) 4506-4512 (2008).
\bibitem{breakjunctions} Reed, M. A., Zhou, C., Muller, C. J., Burgin, T. P. $\&$  Tour, J. M. \textrm{Conductance of a Molecular Junction. }\emph{Science } \textbf{278,} 252-254 (1997).
\bibitem{klein1} Klein, D. L., McEuen, P. L., Bowen, Katari, J. E. B., Roth, R. $\&$ Alivisatos, A. P. \textrm{An approach to electrical studies of single nanocrystals.} \emph{Appl. Phys. Lett} \textbf{68,} 2574-2576 (1996).; Klein, D. L., Roth, R., Lim, A. K. L., Alivisatos, A. P. $\&$ McEuen, P. L. \textrm{A single-electron transistor made from a cadmium selenide nanocrystal.} \emph{Nature (London)} \textbf{389,} 699-701 (1997).

\bibitem{liora} Bitton, L., $\&$ Frydman, A. \textrm{Controllable room-temperature metallic quantum dot. }\emph{Appl. Phys. Lett} \textbf{88,} 113113 (2006).
\bibitem{marcus} Morpurgo, A. F., Marcus, C. M. $\&$ Robinson, D. B. \textrm{Controlled fabrication of metallic electrodes with atomic separation.} \emph{Appl. Phys. Lett.} \textbf{74,} 2084 (1999).
   \bibitem{surface_states} Ray, S. G., Cohen, H., Naaman, R., Liu, H. $\&$ Waldeck, D. H. \textrm{Organization-Induced charge redistribution in self-assembled organic monolayers on gold. } \emph{J. Phys. Chem.} \textbf{B}, 14064-14073 (2005)










\end{references}
\end{document}